# Deep Residual 3D U-Net for Joint Segmentation and Texture Classification of Nodules in Lung[*]

Alexandr Rassadin[1][0000-0002-9492-2895]

[1] xperience.ai
`science@arassadin.com`

**Abstract.** In this work we present a method for lung nodules segmentation, their texture classification and subsequent follow-up recommendation from the CT image of lung. Our method consists of neural network model based on popular U-Net architecture family but modified for the joint nodule segmentation and its texture classification tasks and an ensemble-based model for the follow-up recommendation. This solution was evaluated within the LNDb medical imaging challenge and produced the best nodule segmentation result on the final leaderboard.

## 1 Introduction

Lung cancer is an important disease which can lead to death. Fortunately, nowadays we have early screening procedures for a timely diagnosis of this disease. Early screening with low-dose CT (computed tomography) can reduce mortality by 20% [1]. Unfortunately, global screening of the population would lead to a medical personnel overload. Because of the high patients' flow, doctors lose the ability of steadfast investigation of CT results which can lead to errors in the diagnosis. Nowadays, when artificial intelligence has proven its applicability in many areas of life, shifting routine medical tasks from humans to AI looks like a very desirable option. One of such tasks can be the detection of nodules in the lungs from CT images for follow-up procedures recommendation.

The LNDb challenge [1] consisted of 4 tracks:

- The detection of nodules in lungs from CT images. All nodules from an entire human lung image should be localized.
- The segmentation of the nodules from CT images. Provided with the potential center of the nodule, one should provide accurate voxel-wise binary segmentation of the nodule (if it exists).

---

[*] The authenticated publication is available online at https://doi.org/10.1007/978-3-030-50516-5_37

- The classification of the texture of found nodules. Provided with the center of the potential nodule, one should classify one of the three types of texture: ground glass opacities, part-solid or solid.
- The main challenge track consisted of making a follow-up recommendation based on a CT image according to the 2017 Fleischner society pulmonary nodule guidelines [2].

We participated in all the tracks, except the nodule detection track. The next section describes our approach.

## 2 Method Overview

### 2.1 Nodule Segmentation

We started our experiments with the SSCN [3] U-Net [4] which established itself in a number of 3D segmentation tasks. The advantage of this family of architectures is that it allows the usage of larger batches because of exploiting the sparse nature of the data. Unfortunately, 3D sparse U-Net had a low prediction performance in our setup and our choice was to fall back to the plain 3D convolutions. An analogous setup with 4-stage (i.e. 4 poolings in the encoder part followed by 4 upsamplings in the decoder part) U-Net showed its supremacy over the SSCN counterpart, which determined the direction of further experiments.

The next thing we did was the implementation of residual connections [5], which established itself in many computer vision tasks. Following [6], we replaced the standard ReLU activation with ELU [7] and then the batch normalization [8] with the group normalization [9], which, in combination, gave us a sufficient increase in the segmentation quality. Our final encoder / decoder block is depicted in Fig. 1.

We followed the standard procedure of encoder construction: twice reducing the spatial dimension after every next block while twice increasing the feature dimension (and vice-versa for the decoder). Later we added an additional 5[th] block both to the encoder and decoder, as in [10].

We used the popular attention mechanism CBAM [11] in both the encoder and decoder of our U-Net but adapted it for the 3-dimensional nature of LNDb data. Our experiments show that CBAM leads to a slightly better segmentation quality, which is to be expected, while sacrificing an extra amount of the training time. Here again we see that ELU [7] activation (inside the attention module) provides better results than its ReLU counterpart.

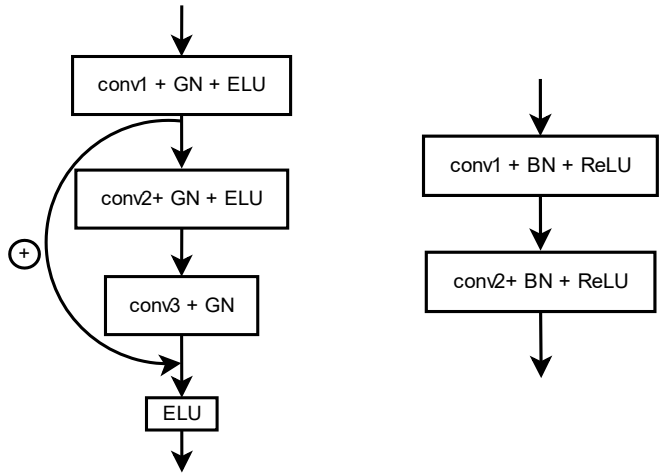

**Fig. 1.** Comparison of the simple U-Net block (right) and our residual block with GroupNorm and ELU (left).

### 2.2 Nodule Texture Classification

As well as for the segmentation track, we started from the SSCN-based recognizer following VGG [12] architecture, but in this case, sparse convolutions failed again. We continued our experiments with the same U-Net encoder which we used for the segmentation but attached a classification head instead of a decoder. The benefit of this approach is that we can start from the already pre-trained for the nodule segmentation weights, which, by our observations, is crucial for training such a classifier. Our classification head starts from the global average pooling followed by two fully connected layers with ELU activation and ends up with the final classification layer with the softmax activation.

Unfortunately, this approach still gave us a low classification accuracy. So, we came up with the idea of training a *joint segmentation and texture classification* network, which is the main contribution of this work. This approach gave us a sufficient boost of the classification accuracy and also increased our nodule segmentation quality because of exploiting so-called multi-task learning.

It is also worth noting that apart from the joint nodule segmentation and texture classification network, we have tried a simple ensemble-based model (namely, Random Forest) upon encoder features from the pre-trained nodule segmentation model. This is a working option but still less accurate than the joint end-to-end model. Also, using a second-stage predictor negatively impacts the performance of the final solution and overcomplicates it.

### 2.3 Joint Nodule Segmentation and Texture Classification

We did some experiments with the texture classification head configuration. Upon our observations, usage of convolutions instead of fully connected layers gives no ad-vantage, neither in terms of classification accuracy nor nodule segmentation quality while the model becomes slightly less robust to the overfitting. For the train/val submission we used a dropout rate 0.6 while for the test set submission a dropout of 0.4 was preferable.

Besides group normalization, we experimented with the switchable normalization [13] and the result was dubious. With the same training procedure, the model with switchable normalization behaves noisier in terms of segmentation and texture classification metrics, slightly worse on average but with few high peaks (see Fig. 2) and it is also more prone to overfit. Such behavior, along with the fact that training with SwitchNorm increases the optimization time, made us fall back to the GroupNorm. We found that the optimal number of groups is 8.

Usage of attention in the encoder and decoder of the joint nodule segmentation and texture classification model has led to the earlier overfeat of the classification head: better segmentation results can be obtained only by sacrificing the texture classification quality. We attempted to overcome this by incorporating some attention mechanism also within the classification head. We tried an approach from [14], again, adapted to the 3D nature of the data, but unfortunately, this approach only decreased the overall quality of the model. So, for the joint model we didn't use any attention mechanism.

The maximum feature size in the first encoder block (see Section 2.1) which fits our hardware (2x Nvidia GTX 1080 Ti) was 40. Unfortunately, the training time of such a model was too high for the limited time budget of the challenge and we decided to use the feature size 32 which still gives us a high enough segmentation quality (see Fig. 3) to provide us with the necessary performance.

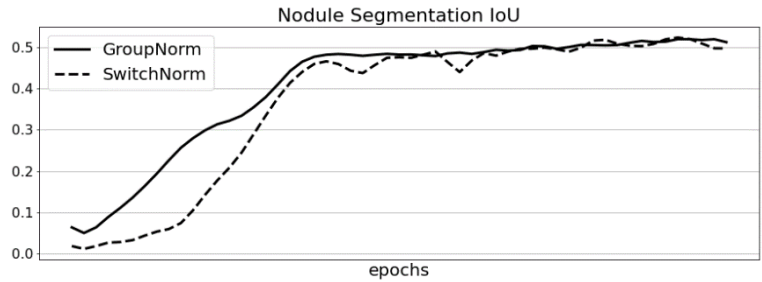

**Fig. 2.** Training curves of two joint nodule segmentation and texture classification networks: one with group normalization and one with switchable normalization.

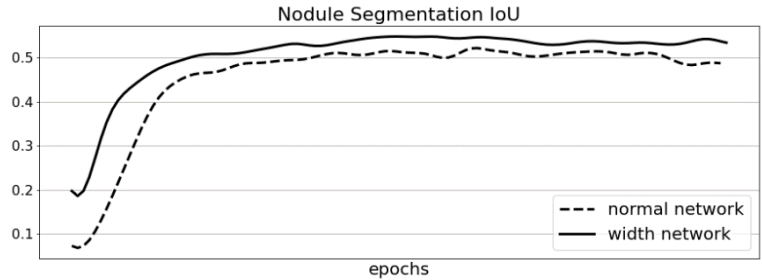

**Fig. 3.** Training curves of normal and width joint nodule segmentation and texture classification networks.

By the nature of the challenge data, participants were able to choose whether to train their models on 5 classes (6 with non-nodule class) or 3 classes (4 with non-nodule class). In our experiments, we clearly observed that the fewer classes to train on – the better final accuracy, so we used a 3-class model in our approach.

### 2.4 Fleischner Classification

As for the main target [2] prediction, we used a relatively simple idea. From the challenge guidelines, we know that the follow-up recommendation can be estimated directly from the nodule annotation considering:

- the number of nodules for the patient (single or multiple),
- their volumes,
- their textures.

This means that information about the number of found nodules, their volumes and textures is enough for a radiologist to make a recommendation. Knowing this, we just encoded this information in a 6-element feature vector as follow:

- The first 3 elements encode the number of nodules of each of the 3 sizes (less than 100 mm$^3$, between 100 and 250 mm$^3$ and more than 250 mm$^3$),
- The last 3 elements encode the number of nodules of each texture type (ground glass opacities, part-solid and solid).

From the predictions of the joint nodule segmentation and texture classification model, we directly know the texture type of the nodule and from the segmentation mask we can compute the nodule size (every nodule has a common spatial resolution).

We first evaluated the prediction capability of such an approach on the ground truth data, using the Random Forest model as a predictor and it showed a remarkable performance – over 90% balanced accuracy. For the leaderboard submissions we just replaced the ground truth segmentation and texture with our own predictions. To overcome the effect of cascade error, we also tried to predict Fleischner target based only on the nodules size (without information of its texture) and surprisingly it had a quite similar prediction capability. Based on these observations, it becomes quite clear that *the crucial factor of the follow-up recommendation is the number of nodules in the lung, which is achieved by the accurate nodule detection or segmentation algorithm.*

The test set of the challenge was extremely noisy due to the false positive nodules in order not to invalidate other tracks targets. Since our team didn't participate in the detection track and the non-nodule filtration is crucial for the main target prediction (because it heavily relies on the information about the number of found nodules in the patient), a strong need arose for some non-nodule filtration mechanism. While submitting the train/val results, this task was assigned to the nodule segmentation network, i.e. a candidate was considered as a non-nodule (false positive) if his volume, based on the predicted nodule segmentation, was nearly zero. We measured the precision of such an approach to non-nodule recognition and it was around 0.64, which, as it turned out, was enough for the slightly noisy train/val data. Looking at the test data, we correctly decided that it would not be sufficient for the highly noisy test data. To solve this problem, we forced to train another auxiliary model, i.e. a separate non-nodule recognizer. For this purpose, we took our joint network without its decoder part, initialized with the best checkpoint, and trained it for the 2-class (nodule / non-nodule) classification problem. Precision of such a model was much higher – 0.78. Incidentally, it was even higher than for a joint model trained for 4 (3 actual classes and 1 non-nodule class) instead of 3 classes. Unfortunately, it turned out that this is still not enough for accurate non-nodule filtration in the test set data which led to the great metrics decrease in the test submission compared to the train/val one (see section 3 for the details).

### 2.5 Model Optimization

Dice loss is default choice nowadays for the training of segmentation models. It worked well in our case too. We experimented with the Generalized Dice overlap loss [15] but it did not give us an improvement. For the classification head we used plain Cross Entropy. We used inversely proportional class weights for Cross Entropy, and it boosted the accuracy, while weighting of the Dice loss didn't provide us with any improvements. Our final loss was an average of the Dice and Cross Entropy, where Cross Entropy was multiplied by 0.2.

As for optimizers, we used very popular Adam optimizer. We also tried recently introduced diffGrad [16] and Adamod [17] but they didn't provide us with any improvements (we didn't perform a hyper-parameter tuning). Comparison of optimizers depicted on Fig. 4. We didn't start optimization of texture classification head (by multiplying Cross Entropy loss by 0) until nodule segmentation achieves 0.45 IoU (intersection over union).

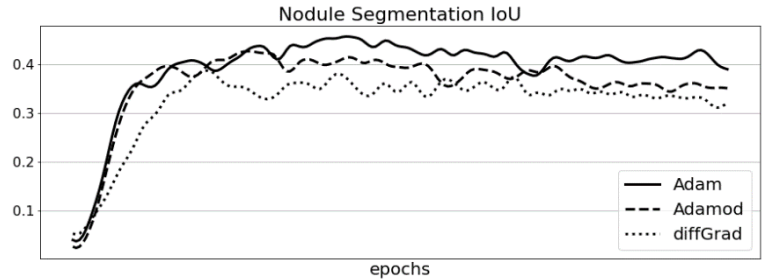

**Fig. 4.** Training curves of the nodule segmentation model for different optimizers.

## 2.6 Data Augmentations

In our work, we used some quite standard augmentations set: random flipping, random rotations by 90 degrees, elastic deformation and noise. We couldn't use rotation for arbitrary angle because it could break the structure of the scan and has padding uncertainty. Our experiments show that augmentations can boost nodule segmentation IoU by 0.05 in average (see Fig. 5).

## 3 Results

### 3.1 Train/val Leaderboard

The organizers provided us with train/val set with 4 predefined folds. Results for the public train/val leaderboard must be submitted using a 4-fold procedure, so we trained 4 joint nodule segmentation and texture classification models. Its predictions were used for the segmentation and texture classification tracks in a straightforward way while for the main target prediction, we first collected the features (volumes and textures) for every nodule, then trained the corresponding predictor for the Fleischner classification (see section 2.4).

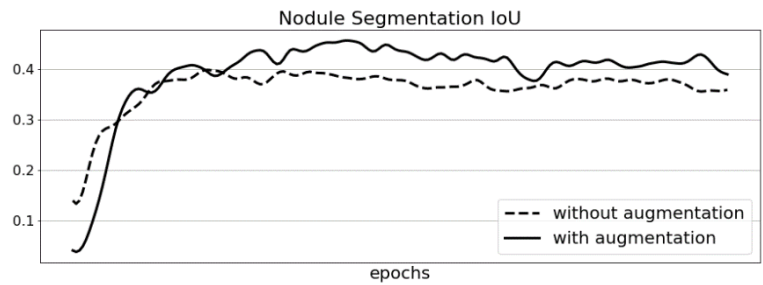

**Fig. 5.** Training curves for two identical nodule segmentation networks: with and without data augmentation.

**Table 1.** Top-5 segmentation results in the train/val leaderboard. Our result highlighted in bold.

| J*     | MAD    | HD     | C*     | Bias     | Std      |
|--------|--------|--------|--------|----------|----------|
| 0.0865 | 0.0827 | 0.4123 | 0.0017 | 16.2162  | 86.1182  |
| 0.4178 | 0.4122 | 2.136  | 0.0671 | 92.7103  | 472.6242 |
| 0.433  | 0.3888 | 2.0493 | 0.0791 | 75.536   | 507.3688 |
| **0.4892** | **0.5668** | **2.4819** | **0.0781** | **103.3177** | **486.8666** |
| 0.3694 | 0.3484 | 1.8991 | 0.1116 | 111.4641 | 719.5786 |

Table 1 summarizes our nodule segmentation result on the train/val leaderboard. Here, $J$ stands for Jaccard index, $MAD$ stands for mean average distance, $HD$ stands for Hausdorff distance, $C$ stands for the Pearson correlation coefficient, $Bias$ stands for the mean absolute difference, $Std$ stands for the standard deviation, symbol * stands for the inversion of the metric, e.g. $J*$ means $1 - J$. The final score in the leaderboard was calculated as an average of all six metrics, which were preliminarily normalized by the maximum value over all the submissions in the leaderboard, for each metric separately. See the LNDb challenge evaluation page [18] for a detailed description of the evaluation metrics.

Our Fleischner classification score is 0.5281 Fleiss-Cohen weighted kappa [19], which is the third best result in the leaderboard (after 0.603 and 0.5323 kappa).

### 3.2    Test Leaderboard

We used 70% of the train/val data for training our joint nodule segmentation and texture classification model while the remaining 30% were used for validation and also for the training of our main target predictor.

Nodule segmentation and its texture classification procedures were the same as for the train/val submission – results were obtained in a straightforward way from the joint model.

Additionally, for every sample, we predicted whether or not it is a nodule using our non-nodule recognition model (see Section 2.4) and saved this information to make a later prediction of the main target. Then we took the remaining 30% of the train/val set, which was not used for training the joint model, and collected the segmentation, texture classification and non-nodule recognition results for this data. From this prediction we formed a sampling for training a Random Forest predictor of the main target (see Section 2.4). Finally, this model was used for the prediction of the main target of the test set.

Table 2 summarizes our nodule segmentation results on the test leaderboard. See the LNDb challenge evaluation page [18] for a detailed description of the evaluation metrics.

Our Fleischner classification score is -0.0229 Fleiss-Cohen weighted kappa [19]. We explained the reasons of such a poor result and the significant difference with the train/val submission in Section 2.4 in detail.

**Table 2.** Top-3 segmentation results in the test leaderboard. Our result highlighted in bold.

| J* | MAD | HD | C* | Bias | Std |
|---|---|---|---|---|---|
| **0.4779** | **0.4203** | **2.0275** | **0.055** | **44.2826** | **86.3227** |
| 0.468 | 0.4686 | 2.1371 | 0.081 | 40.701 | 98.741 |
| 0.4447 | 0.4115 | 2.0618 | 0.1452 | 41.4341 | 129.47 |

## 4 Conclusions

In this paper we described a solution for lung nodules segmentation, their texture classification and a consequent follow-up recommendation for the patient. Our approach consists of a joint nodule segmentation and texture classification neural network, which is essentially a deep residual U-Net [4] with batch normalization [8] replaced by a group normalization [9] and ReLU replaced by ELU [7]. For the patient's follow-up recommendation [2], we used an ensemble-based model. We evaluated our approach by participating in the LNDb challenge [1] and took the first place in the segmentation track with a result of 0.5221 IoU. Our approach is simple yet effective and can potentially be used in real diagnostic systems reducing the routine workload on medical personnel, which clearly defines the direction of our future work.

## Acknowledgments

We are grateful to xperience.ai for support of the research and Andrey Savchenko for his assistance in preparation of this paper.

## References

1. Pedrosa, J., Aresta, G., Ferreira, C., Rodrigues, M., Leitão, P., Carvalho, A. S., ... & Campilho, A. (2019). LNDb: A Lung Nodule Database on Computed Tomography. arXiv preprint arXiv:1911.08434 (2019).
2. MacMahon, H., Naidich, D., Goo, J., Lee, K., Leung, A., Mayo, J., Mehta, A., Ohno, Y., ... & Bankier, A.: Guidelines for Management of Incidental Pulmonary Nodules Detected on CT Images: From the Fleischner Society 2017. Radiology 284(1), 228-243 (2017).
3. Graham, B., Maaten, L.: Submanifold Sparse Convolutional Networks. arXiv preprint arXiv:1706.01307 (2017).
4. Ronneberger, O., Fischer, P., Brox, T.: U-Net: Convolutional Networks for Biomedical Image Segmentation. arXiv preprint arXiv:1505.04597 (2015).
5. He, K., Zhang, X., Ren, S., Sun, J.: Deep Residual Learning for Image Recognition. arXiv preprint arXiv:1512.03385 (2015).
6. Lee, K., Zung, J., Li, P., Jain, V., Seung, H. S.: Superhuman Accuracy on the SNEMI3D Connectomics Challenge. arXiv preprint arXiv:1706.00120 (2017).
7. Clevert, D. A., Unterthiner, T., Hochreiter, S.: Fast and Accurate Deep Network Learningby Exponential Linear Units (ELUs). arXiv preprint arXiv:1511.07289 (2015).

8. Ioffe, S., Szegedy, C.: Batch Normalization: Accelerating Deep Network Training by Reducing Internal Covariate Shift. arXiv preprint arXiv:1502.03167 (2015).
9. Wu, Y., He, K.: Group Normalization. arXiv preprint arXiv:1803.08494 (2018).
10. Wolny, A., Cerrone, L., … & Kreshuk, A.: Accurate and Versatile 3D Segmentation of Plant Tissues at Cellular Resolution. bioRxiv preprint doi:10.1101/2020.01.17.910562 (2020).
11. Woo, S., Park, J., Lee, J. Y., Kweon, I. S.: CBAM: Convolutional Block Attention Module. arXiv preprint arXiv:1807.06521 (2018).
12. Simonyan, K., Zisserman, A.: Very Deep Convolutional Networks for Large-Scale Image Recognition. arXiv preprint arXiv:1409.1556 (2014).
13. Luo, P., Ren, J., Peng, Z., Zhang, R., Li, J.: Differentiable Learning-to-Normalize via Switchable Normalization. arXiv preprint arXiv:1806.10779 (2018).
14. Lin, T. Y., RoyChowdhury, A., Maji, S.: Bilinear CNNs for Fine-grained Visual Recognition. arXiv preprint arXiv:1504.07889 (2015).
15. Sudre, C., Li, W., Vercauteren, T., Ourselin, S., Cardoso, M.: Generalised Dice overlap as a deep learning loss function for highly unbalanced segmentations. arXiv preprint arXiv: 1707.03237 (2017).
16. Dubey, S., Chakraborty, S., Roy, S., … & Chaudhuri, B.: diffGrad: An Optimization Method for Convolutional Neural Networks. arXiv preprint arXiv:1909.11015 (2019).
17. Ding, J., Ren, X., Luo, R., Sun, X.: An Adaptive and Momental Bound Method for Stochastic Learning. arXiv preprint arXiv:1910.12249 (2019).
18. LNDb challenge evaluation page, https://lndb.grand-challenge.org/Evaluation/, last accessed 2020/02/08.
19. Spitzer, R., Cohen, J., Fleiss, J., Endicott, J.: Quantification of Agreement in Psychiatric Diagnosis: A New Approach. Archives of General Psychiatry 17(1), 83-87 (1967).